\documentclass[12pt]{article}
\usepackage[dvips]{color}
\usepackage{epsfig}
\usepackage{amsmath}
\usepackage{graphicx}
\def\Box{\hbox{$\rlap{$\sqcup$}\sqcap$}}
\begin{document}
\setcounter{page}{1}

\pagestyle{plain} \vspace{1cm}

\begin{center}
\Large{\bf Phantom Behavior Bounce with Tachyon and Non-minimal Derivative Coupling}\\
\small \vspace{1cm} {\bf A. Banijamali $^{a}$
\footnote{a.banijamali@nit.ac.ir}} and {\bf B. Fazlpour $^{b}$
\footnote{b.fazlpour@umz.ac.ir}}\\
\vspace{0.5cm}  $^{a}$ {\it Department of Basic Sciences, Babol
University of Technology, Babol, Iran\\} \vspace{0.5cm} $^{b}$ {\it
Young researchers club, Ayatollah Amoli branch, Islamic azad
university, Amol, Iran\\}
\end{center}
\vspace{1.5cm}
\begin{abstract}
The bouncing cosmology provides a successful solution of the
cosmological singularity problem. In this paper, we study the
bouncing behavior of a single scalar field model with tachyon field
non-minimally coupled to itself, its derivative and to the
curvature. By utilizing the numerical calculations we will show that
the bouncing solution can appear in the universe dominated by such a
quintom matter with equation of state crossing the phantom divide
line. We also investigate the classical stability of our model using
the phase velocity of the homogeneous perturbations of
the tachyon scalar field.\\

{\bf PACS numbers:} 98.80.cq, 98.80.-k, 04.50.kd\\
{\bf Keywords:} Bouncing universe; Non-minimal derivative coupling;
Tachyon field; Stability.\\

\end{abstract}
\newpage
\section{Introduction}
Cosmological observations have revealed that our universe has
experienced a phase of accelerated expansion in the early-time known
as inflation [1]. Although inflation has great successes such as
solving the flatness, horizon and homogeneity problems [2], but it
suffers from the initial singularity problem and hence cannot give
complete descriptions of the early universe.\\
A successful solution of the cosmological singularity problem may be
provided by non-singular bouncing cosmology which has been proposed
a long time ago [3]. In order to circumvent the Big-Bang singularity
problem one needs to choose one of the following approaches: 1)
modified gravity [4], 2) using a kind of matter field that violates
the null energy condition (see [5] for review). Through the modified
gravity approach various scenarios have been constructed such as the
brane world models [6], Pre-Big-Bang [7] and the Ekpyrotic models
[8], $f(T)$ gravity in which gravity is described by an arbitrary
function of the torsion scalar [9], higher derivative gravity
actions [10], non-relativistic gravitational actions [11, 12] or
loop quantum cosmology [13]. Bouncing cosmologies can also be
inspired by string theory when the string gas cosmology
[14, 15] embed in a bouncing universe [16].\\
Furthermore, Bouncing cosmology for a homogeneous and isotropic
universe can lead to the matter bounce scenario. Non-conventional
fluids [17, 18], non-minimally coupled scalar fields to gravity [19]
and ghost condensates [20-22] are belong to this sort of bouncing
cosmology. The simplest possibility to obtain a bounce is a massive
scalar field in the presence of a positive scalar curvature [23] and
the other type of the matter bounce cosmology is
the quintom bounce model [24, 25].\\
Moreover, various observational data such as SNe Ia Gold data set
[26] confirmed that the effective equation of state (EoS) parameter
$\omega$ (the ratio of the effective pressure of the universe to the
effective energy density of it) cross $-1$, namely, the phantom
divide line, currently or in the past.\\
The scenarios with phantom behavior are designed to understand the
nature of dark energy with $\omega$ across $-1$. To realize a viable
scenario with phantom behavior one needs to introduce extra degree
of freedom to the conventional theory with a single fluid or a
single scalar field. The fact that the phantom behavior can be
achieved without ghosts in scalar-tensor gravity was first shown in
[27]. Also, quintom scenario which realizes the phantom behavior
first introduced in Ref. [28] with two scalar fields (quintessence
and phantom). Another attempts for constructing a model that shows
phantom behavior are as follows: scalar field model with non-linear
kinetic terms [29] or a non-linear higher-derivative one [30],
braneworld models [31], phantom coupled to dark matter with an
appropriate coupling [32], string inspired models [33], non-local
gravity [34], modified gravity models [35] and also non-minimally
coupled scalar field models in which scalar field couples with
scalar curvature, Gauss-Bonnet invariant or modified $f(R)$ gravity
[36-38] (for a detailed review, see [39]). Crossing of the phantom
divide can also be realized with single imperfect fluid [40] or by a
constrained
single degree of freedom dust like fluids [41].\\
Moreover, non-minimal couplings are generated by quantum corrections
to the scalar field theory and they are essential for the
renormalizability of the scalar field theory in curved space (see
[42] and references therein). One can extend the non-minimally
coupled scalar tensor theories, allowing for non-minimal coupling
between the derivatives of the scalar fields and the curvature [43].
A model with non-minimal derivative coupling was proposed in [43-45]
and interesting cosmological behaviors of such a model in
inflationary cosmology [46], quintessence and phantom cosmology [47,
48], asymptotic solutions and restrictions on the coupling parameter
[49] have been widely studied in the literature. General non-minimal
coupling of a scalar field and kinetic term to the curvature as a
source of dark energy has been analyzed in [50]. Also, non-minimal
coupling of modified gravity and kinetic part of Lagrangian of a
massless scalar field has been investigated in [51]. It has been
shown that inflation and late-time cosmic acceleration of the
universe can be realized in such a model. \\
In this paper we consider a model with an explicit coupling between
the scalar field, the derivative of the scalar field and the
curvature and study the bouncing solution in such a model. We are
interested in our analysis to the case of tachyon scalar field. The
tachyon field in the world volume theory of the open string
stretched between a D-brane and an anti-D-brane or a non-BPS D-brane
plays the role of scalar field in the context of string theory [52].
What distinguishes the tachyon Lagrangian from the standard
Klein-Gordan form for scalar field is that the tachyon action has a
non-standard type namely, Dirac-Born-Infeld form [53]. Our
motivation for investigating a model with non-minimal derivative
coupling and tachyon scalar field is coming from a fundamental
theory such as string/superstring theory and it may provide a
possible approach to quantum gravity from a perturbative point of view [54-56].\\
An outline of the present work is as follows: In section 2 we
introduce our model in which the tachyon field plays the role of
scalar field and the non-minimal coupling between scalar field, the
derivative of scalar field and Einstein tensor is also present in
the action. Then, we derive the field equations as well as the
energy density and pressure of the model in order to study the
bouncing behavior of our model. We obtain the condition required for
bouncing solution of our model and then we show that such a
condition can be satisfied numerically. In Section 3 we discuss the
classical stability of our model. Section 4 is devoted to our conclusions.\\

\section{Bouncing behavior of non-minimal derivative coupling of tachyon gravity}
In this section we review the field equations of tachyon gravity
with non-minimal derivative coupling and provide the cosmological
equations in a Friedmann-Robertson-Walker (FRW) universe. Then, we
obtain the necessary conditions required for a successful bounce in
such a model. Our starting point is the following Born-Infeld type
action for tachyon field with non-minimal derivative coupling and
also with itself,
\begin{equation}
S=\int d^{4}x
\sqrt{-g}\Big[\frac{1}{2\kappa^{2}}(R-2\Lambda)-V(\phi)\sqrt{1+g^{\mu\nu}\partial_{\mu}\phi\partial_{\nu}\phi}+\xi
f(\phi)G_{\mu\nu}\partial^{\mu}\phi\partial^{\nu}\phi\Big],
\end{equation}
where $\kappa^{2} = 8\pi G = \frac{1}{M_{Pl}^{2}}$ while $G$ is a
bare gravitational constant and $M_{Pl}$ is a reduced Planck mass,
$\Lambda$ is the cosmological constant and $V(\phi)$ is the tachyon
potential which is bounded and reaching its minimum asymptotically.
$f(\phi)$ is a general function of the tachyon field $\phi$ and
$\xi$ is coupling constant. The models of kind (1) with non-minimal
coupling between derivatives of a scalar field and curvature are the
extension of scalar-tensor theories. Such a non-minimal coupling may
appear in some Kaluza-Klein theories [57, 58]. In Ref. [43],
Amendola has considered a model with non-minimal coupling between
derivative of scalar field and the Ricci scalar, $\xi
R\partial_{\mu}\phi\partial^{\mu}\phi$, and by using generalized
slow-roll approximations, he has obtained some
inflationary solutions of this model.\\
A general model containing two derivative coupling terms $\xi_{1} R
\partial_{\mu}\phi \partial^{\mu}\phi$ and $\xi_{2} R_{\mu\nu} \partial^{\mu}\phi
\partial^{\nu}\phi$, has been studied in [44, 45]. It was shown in [47]
that field equations of this theory are of third order in
$g_{\mu\nu}$ and $\phi$, but in the special case where
$-2\xi_{1}=\xi_{2}=\xi$ the order of equations are reduced to the
second order. This particular choice of $\xi_{1}$ and $\xi_{2}$
leads to the non-minimal coupling between derivative of scalar field
and the Einstein tensor, $\xi G_{\mu\nu}
\partial^{\mu}\phi\partial^{\nu}\phi$. Sushkov in [47] has obtained
the exact cosmological solutions of this theory and he has concluded
that such a model is able to explain a quasi-de sitter phase as
well as an exit from it without any fine-tuned potential.\\
Let us now present the cosmological equations. Variation of action
(1) with respect to the metric tensor $g_{\mu\nu}$ gives,
\begin{equation}
R_{\mu\nu}-\frac{1}{2}g_{\mu\nu}R+g_{\mu\nu}\Lambda=\kappa^{2}\big(T_{\mu\nu}+\xi
T'_{\mu\nu}\big),
\end{equation}
where
\begin{equation}
T_{\mu\nu}=V(\phi)\Big(\frac{\nabla_{\mu}\phi
\nabla_{\nu}\phi}{u}-g_{\mu\nu}u\Big),
\end{equation}
and
$$T'_{\mu\nu}=R\big(\nabla_{\mu}\phi\nabla_{\nu}\phi\big)-4\nabla_{\gamma}\phi\nabla_{(\mu}\phi
R^{\gamma}_{\nu)}+G_{\mu\nu}\big(\nabla\phi\big)^{2}-2R_{\mu\nu\gamma\lambda}\nabla^{\gamma}
\phi\nabla^{\lambda}\phi-2\nabla_{\mu}\nabla^{\gamma}\phi\nabla_{\nu}\nabla_{\gamma}\phi$$
\begin{equation}
+2\nabla_{\mu}
\nabla_{\nu}\phi\Box\phi+g_{\mu\nu}\Big(\nabla^{\gamma}\nabla^{\lambda}\phi\nabla_{\gamma}\nabla_{\lambda}
\phi-\big(\Box\phi\big)^{2}+2R^{\gamma\lambda}\nabla_{\gamma}\phi\nabla_{\lambda}\phi\Big),
\end{equation}
where $u=\sqrt{1+\nabla_{\mu}\phi\nabla^{\mu}\phi}$.\\
One can obtain the scalar field equation of motion by varying (1)
with respect to $\phi$,
\begin{equation}
\nabla_{\mu}\Big(\frac{V(\phi)\nabla^{\mu}\phi}{u}\Big)-\frac{dV(\phi)}{d\phi}u-\xi
f(\phi) G^{\mu\nu}\nabla_{\mu}\nabla_{\nu}\phi+\xi
\frac{df(\phi)}{d\phi}G_{\mu\nu}\partial^{\mu}\phi\partial^{\nu}\phi=0.
\end{equation}
For a flat FRW metric,
\begin{eqnarray}
ds^{2}=-dt^{2}+a^{2}(t)(dr^{2}+r^{2}d\Omega^{2}),
\end{eqnarray}
the  components of the Ricci tensor $R_{\mu\nu}$ and the Ricci
scalar $R$ are given by
\begin{equation}
R_{00}=-3\big(\dot{H}+H^{2}\big),\,\,R_{ij}=a^{2}(t)\big(\dot{H}+3H^{2}\big)\delta_{ij},\,\,
R=6\big(\dot{H}+2H^{2}\big),
\end{equation}
where $H=\frac{\dot{a}(t)}{a(t)}$ is the Hubble parameter and $a(t)$
is the scale factor. The scalar field equation of motion for a
homogeneous $\phi$ in FRW background (6) takes the following form
$$\frac{\ddot{\phi}}{1-\dot{\phi}^{2}}+3H\dot{\phi}+\frac{1}{V(\phi)}\frac{dV}{d\phi}$$
\begin{equation}
+\frac{\sqrt{1-\dot{\phi}^{2}}}{V(\phi)} \Bigg(3\xi
H^{2}\Big(2f(\phi)\ddot{\phi}+\frac{df}{d\phi}\dot{\phi}^{2}\Big)+18\xi
H^{3}f(\phi)\dot{\phi}+12\xi H \dot{H} f(\phi)\dot{\phi}\Bigg)=0.
\end{equation}
From equation (2) the energy density and pressure are as follows
respectively,
\begin{equation}
\rho=\frac{V(\phi)}{\sqrt{1-\dot{\phi}^{2}}}+9\xi
H^{2}f(\phi)\dot{\phi}^{2}+\frac{\Lambda}{\kappa^{2}},
\end{equation}
and
\begin{equation}
P=-V(\phi)\sqrt{1-\dot{\phi}^{2}}-\xi\big(3H^{2}+2\dot{H}\big)f(\phi)\dot{\phi}^{2}-2\xi
H\Big(2f(\phi)\dot{\phi}\ddot{\phi}+\frac{df}{d\phi}\dot{\phi}^{3}\Big)-\frac{\Lambda}{\kappa^{2}}.
\end{equation}
Friedmann equation is also as follows,
\begin{equation}
H^{2}=\frac{\kappa^{2}}{3}\Big(\frac{V(\phi)}{\sqrt{1-\dot{\phi}^{2}}}+9\xi
H^{2}f(\phi)\dot{\phi}^{2}\Big)+\frac{\Lambda}{3}.
\end{equation}
Moreover, a bouncing universe starts with an initial contraction to
a non-vanishing minimal radius, then subsequent an expanding phase
[24]. During the contracting phase, the scale factor $a(t)$ is
decreasing, i.e., $\dot{a} < 0$, and in the expanding phase we have
$\dot{a}> 0$. At the bouncing point, $\dot{a} = 0$, and around this
point $\ddot{a} > 0$ for a period of time. Equivalently in the
bouncing cosmology the Hubble parameter $H$ runs across zero from $H
< 0$ to $H > 0$ and $H = 0$ at the bouncing point. A successful
bounce requires around this point,
\begin{equation}
\dot{H}=-\frac{\kappa^{2}}{2}(\rho+P)=-\frac{\kappa^{2}}{2}\rho(1+\omega)>0.
\end{equation}
As it is clear from equation (12), a successful non-singular
bouncing cosmology violates the null energy condition \footnote{The
NEC for a null vector $k^{\mu}$, implies
$T_{\mu\nu}k^{\mu}k^{\nu}\geq0$ which for a perfect fluid reads
$\rho+P\geq0$.}(NEC) around the bouncing point. In addition, in
order to achieve the hot Big-Bang era after the bouncing, the EoS of
the universe, $\omega$, must cross the so
called phantom divide line $\omega=-1$.\\
A salient feature of our model (1) is that it realizes the phantom
divide line crossing [59]. In Ref. [59] we have presented the
conditions required for such a crossing and numerically shown that
this model realizes crossing of the $\omega=-1$ for a special case
of the Hubble parameter $H=\frac{h_{0}}{t}$ with constant $h_{0}$.
In the present work we show that crossing over $-1$ can take place
in our model for a general Hubble parameter obtained from
equation (11).\\
To study the bouncing behavior of the present model we use the
equation (12). From equations (9), (10) and (12) one can see that a
successful bounce requires:
\begin{equation}
\frac{V(\phi) \dot{\phi}^{2}}{\sqrt{1-\dot{\phi}^{2}}}+6\xi
H^{2}f(\phi)\dot{\phi}^{2}-2\xi\dot{H}f(\phi)\dot{\phi}^{2}-2\xi
H\Big(2f(\phi)\dot{\phi}\ddot{\phi}+\frac{df}{d\phi}\dot{\phi}^{3}\Big)<0.
\end{equation}
We will show below that our model can easily fulfilled the above
bouncing condition. Note that in the numerical study on the bouncing
solution, we have used the following state for Hubble parameter
obtained from equation (11):
\begin{equation}
H^{2}=\frac{1}{\big(\frac{3}{k^{2}}-9\xi
f(\phi)\dot{\phi}^{2}\big)}\Big(\frac{V(\phi)}{\sqrt{1-\dot{\phi}^{2}}}+\frac{\Lambda}{\kappa^{2}}\Big).
\end{equation}
In figures 1 and 2, we show the bouncing solution for two different
potentials. Figure 1 shows such a bouncing solution for
$V(\phi)=V_{0}e^{-\alpha \phi^{2}}$ with constant $\alpha$. One can
see from this figure that the model predicts crossing of the
$\omega=-1$, which gives rise to a possible inflationary phase after
the bounce. Also, the Hubble parameter evolution as a function of
cosmic time has been shown in figure 1. One can read from this
figure that the Hubble parameter $H$ running across zero at $t=0$
which have choosed it, as the bouncing point. In figure 2, we
consider the model with potential $V(\phi)=\frac{V_{0}}{\phi^{2}}$
and then show another example of the bouncing behavior of the model.
This figure show the crossing over $-1$ for the EoS again. In the
numerical calculations we have used the function $f(\phi)=b\phi^{n}$
with constants $b$ and $n$.\\

\begin{figure}[htp]
\begin{center}
\includegraphics{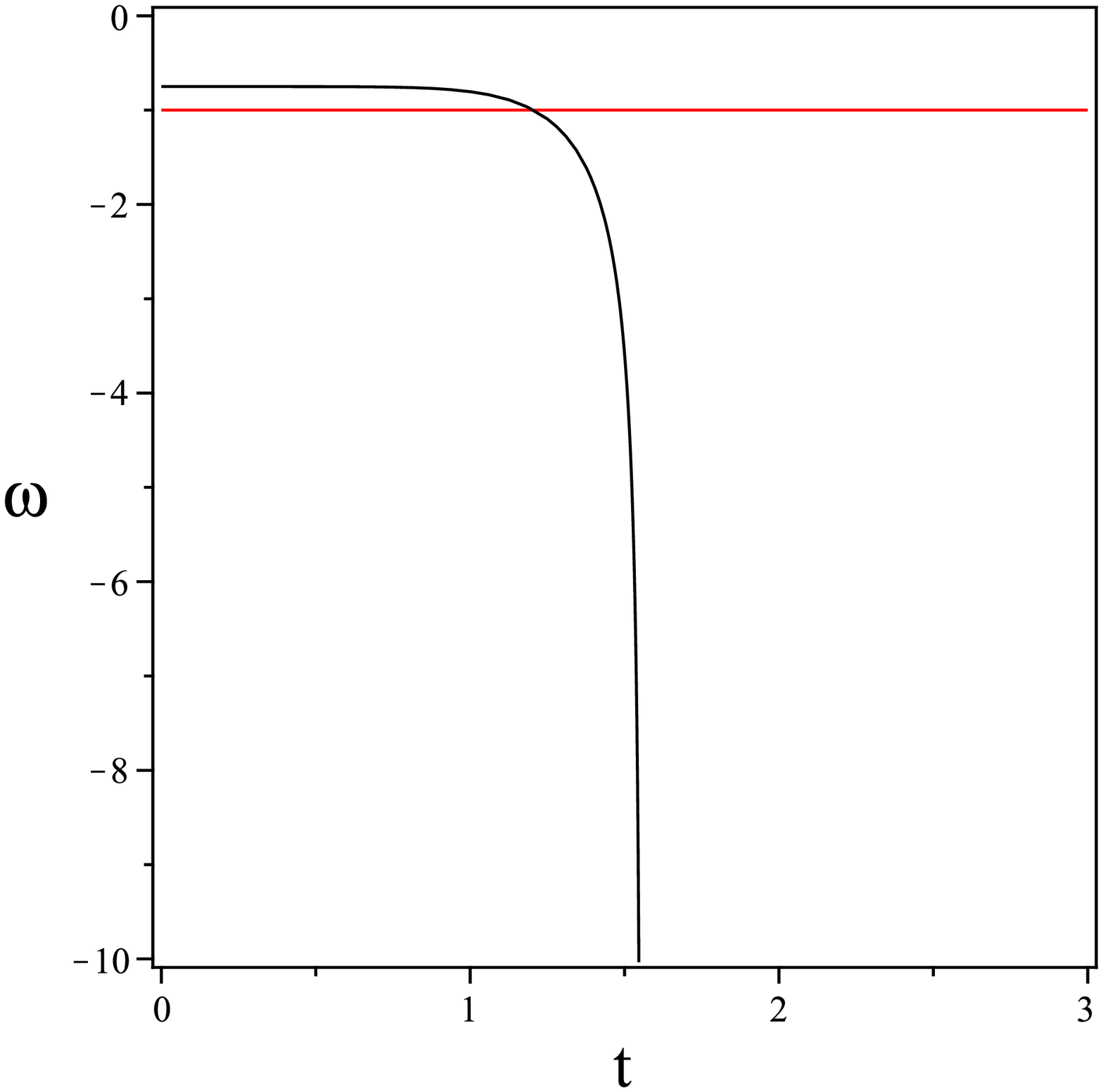} \vspace{7cm}\includegraphics{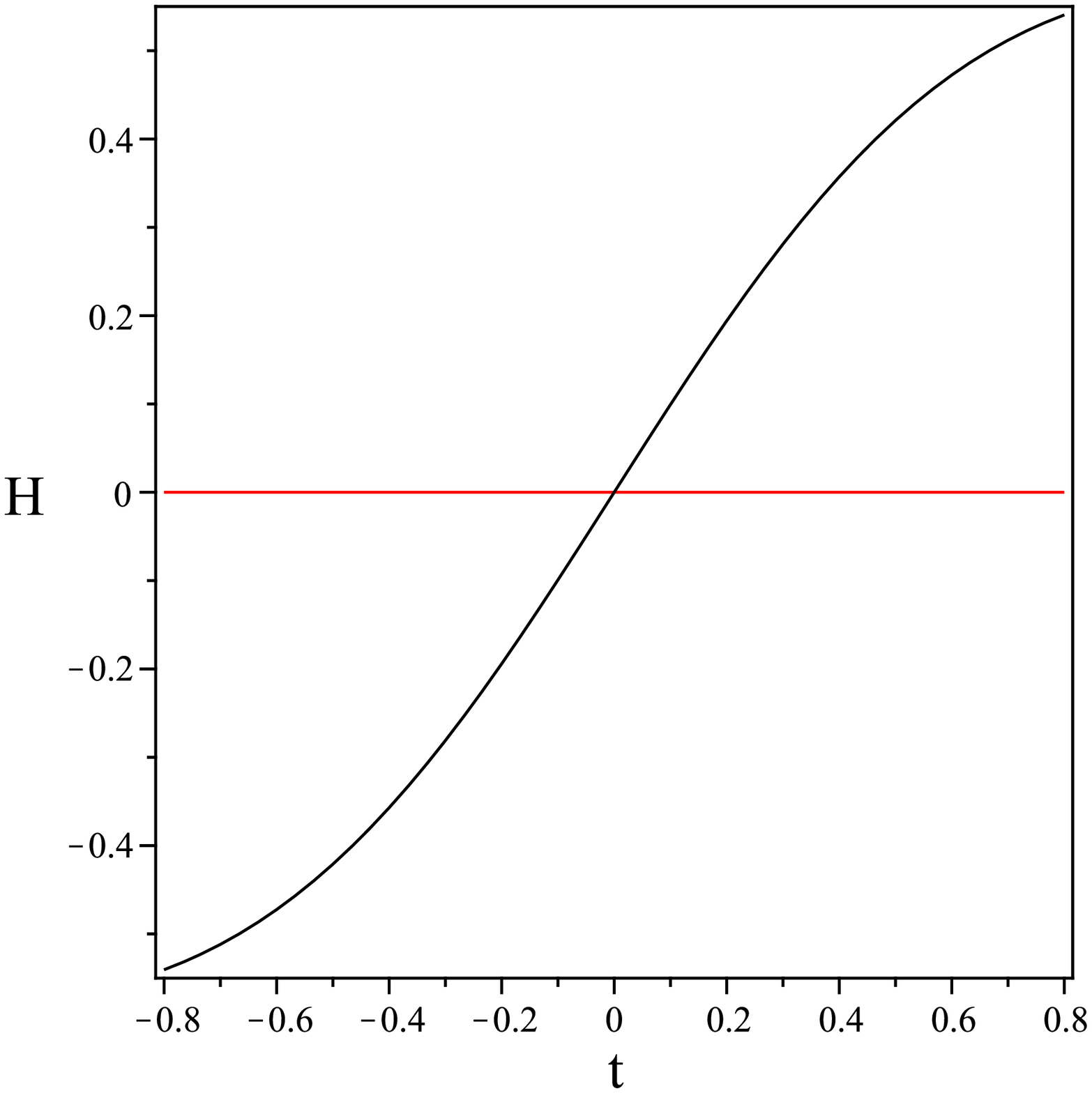}
\end{center}
 \caption{\small { Plots of the evolution of the
EoS and the Hubble parameter versus $t$ for the potential
 $V(\phi)=V_{0}e^{-\alpha \phi^{2}}$, $\phi=\phi_{0}t$, $f(\phi)=b\phi^{n}$ and negative cosmological constant
 $\Lambda=-0.05$, (with $\xi=10$, $b=1$,
 $n=8$, $V_{0}=0.04$, $\phi_{0}=0.6$ and $\alpha=5$).}}
\end{figure}

\begin{figure}[htp]
\begin{center}
\includegraphics{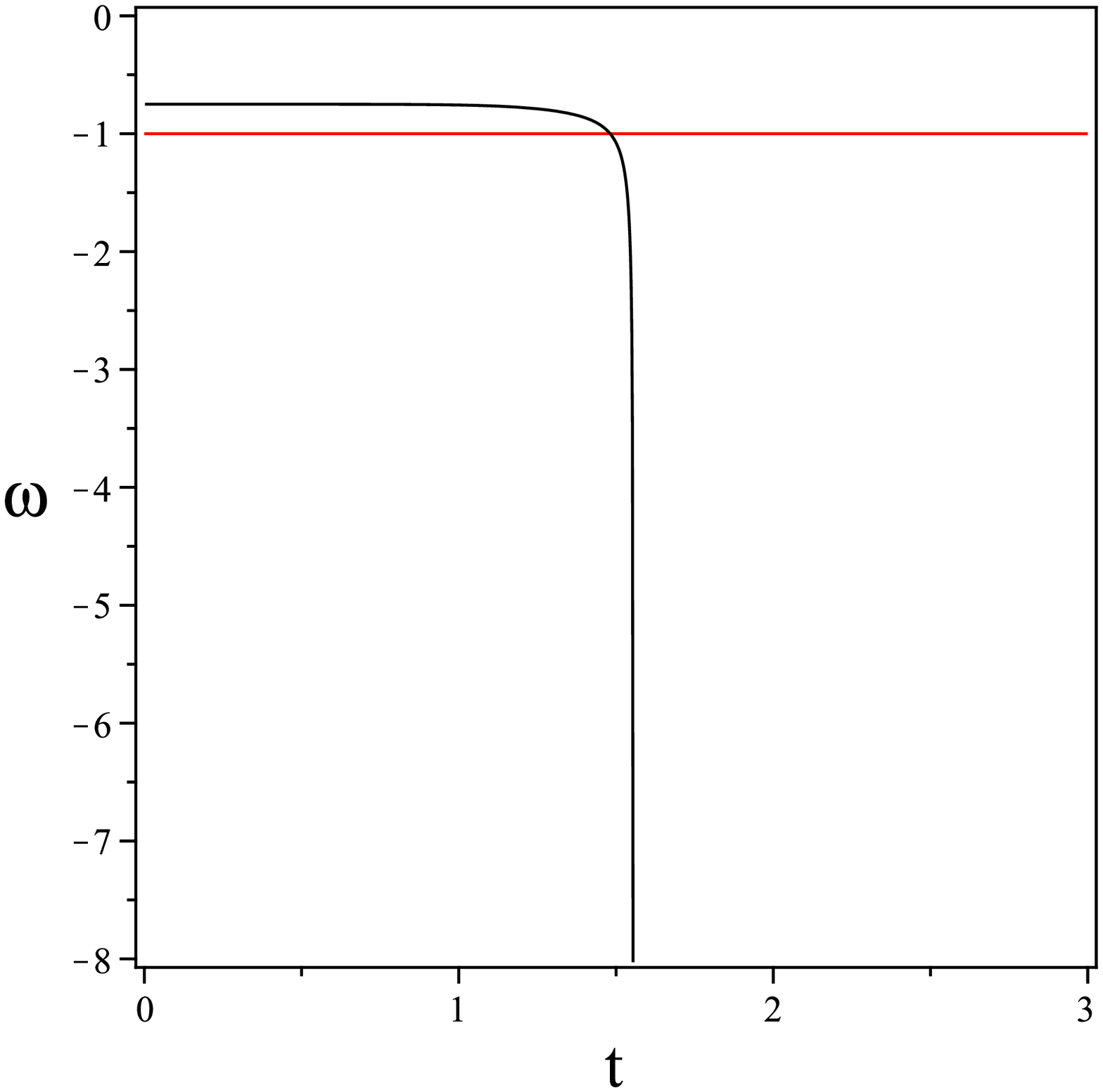} \vspace{7cm}\includegraphics{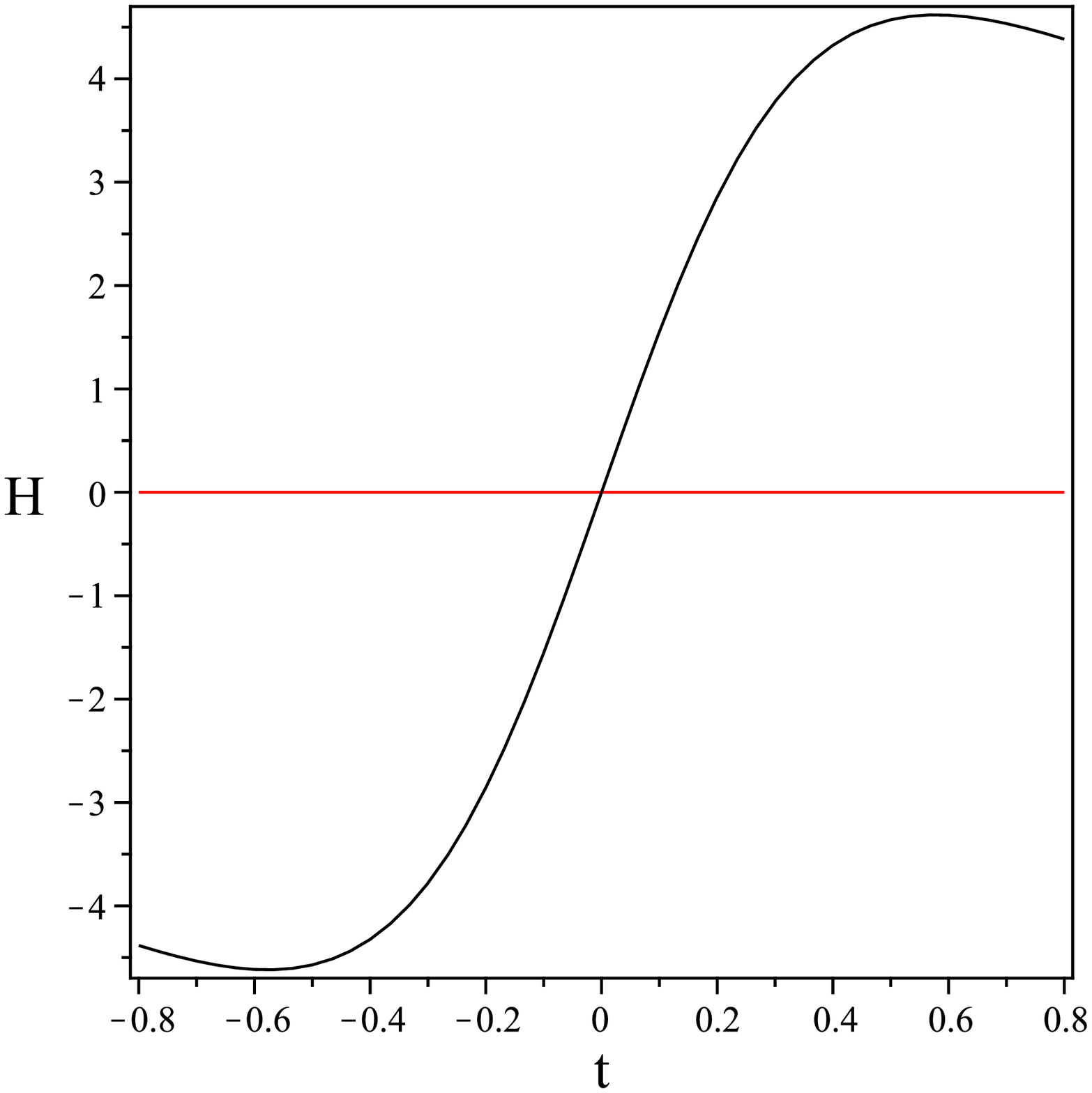}
\end{center}
 \caption{\small { Plots of the evolution of the
EoS and the Hubble parameter versus $t$ for the potential
 $V(\phi)=\frac{V_{0}}{\phi^{2}}$, $\phi=\phi_{0}t$, $f(\phi)=b\phi^{n}$ and $\Lambda=-0.05$, (with $\xi=10$, $b=1$,
 $n=8$, $V_{0}=0.04$ and $\phi_{0}=0.6$).}}
\end{figure}

\newpage
\section{Stability of the model}
Now we discuss on the stability of the model. The sound speed,
$C_{s}^{2}$, is the function appearing before spatial gradients in
the scalar field equation of motion and for perturbations around
homogeneous solutions, it is a function of time. The sound speed
express the phase velocity of the inhomogeneous perturbations of the
tachyon field [60]. The absence of gradient instabilities requires
$C_{s}^{2}\geq0$, where
\begin{equation}
C_{s}^{2}=\frac{P'}{\rho'},
\end{equation}
where a prim denotes derivative with respect to $\dot{\phi}^{2}$. By
using equations (9), (10) and (14), the sound speed parameter reads,
$$C_{s}^{2}=\frac{2k^{2}}{3\big(\frac{3}{k^{2}}-27\xi
f(\phi)\dot{\phi}^{2}+18\xi f(\phi)\big)}\Bigg[\frac{3\xi
f(\phi)\dot{\phi}}{\big(\frac{3}{k^{2}}-9\xi
f(\phi)\dot{\phi}^{2}\big)^{\frac{1}{2}}(1-\dot{\phi}^{2})^{\frac{1}{4}}}\Bigg
(6\xi\dot{\phi}^{2}(1-\dot{\phi}^{2})\frac{df}{d\phi}$$
$$+\frac{\ddot{\phi}}{k^{2}}+\frac{9}{4}\xi\dot{\phi}^{4}
\frac{df}{d\phi}+\frac{7\dot{\phi}^{2}\ddot{\phi}}{4k^{2}(1-\dot{\phi}^{2})}-\frac{21\xi
f(\phi)\dot{\phi}^{4}\ddot{\phi}}{4(1-\dot{\phi}^{2})}+6\xi
f(\phi)\ddot{\phi}(1-\dot{\phi}^{2})-\frac{3}{2}\xi
f(\phi)\dot{\phi}^{2}\ddot{\phi}\Bigg)$$
$$+\frac{135\xi^{2}
f^{2}(\phi)\dot{\phi}^{3}}{2\big(\frac{3}{k^{2}}-9\xi
f(\phi)\dot{\phi}^{2}\big)^{\frac{3}{2}}(1-\dot{\phi}^{2})^{\frac{1}{4}}}\Bigg
(3\xi\dot{\phi}^{2}(1-\dot{\phi}^{2})\frac{df}{d\phi}+\frac{\ddot{\phi}}{k^{2}}-3\xi
f(\phi)\dot{\phi}\ddot{\phi}$$
$$+6\xi
f(\phi)\ddot{\phi}(1-\dot{\phi}^{2})\Bigg)+27\xi^{2}f^{2}(\phi)\dot{\phi}^{2}
(1-\dot{\phi}^{2})+3\xi f(\phi)\big(\frac{3}{k^{2}}-9\xi
f(\phi)\dot{\phi}^{2}\big)(1-\frac{1}{2}\dot{\phi}^{2})$$
$$-\frac{1}{2\sqrt{(1-\dot{\phi}^{2})}}+\frac{1}{\sqrt{V(\phi)}}\Bigg(\xi\dot{\phi}
\big(\frac{3}{k^{2}}-9\xi
f(\phi)\dot{\phi}^{2}\big)^{\frac{3}{2}}(1-\dot{\phi}^{2})^{\frac{1}{4}}\big(\frac{1}{2}
\dot{\phi}^{2}\frac{df}{d\phi}+f(\phi)\ddot{\phi}+2(1-\dot{\phi}^{2})\frac{df}{d\phi}\big)$$
$$+9\xi^{2}f(\phi)\dot{\phi}\big(\frac{3}{k^{2}}-9\xi
f(\phi)\dot{\phi}^{2}\big)^{\frac{1}{2}}(1-\dot{\phi}^{2})^{\frac{5}{4}}\big(
\dot{\phi}^{2}\frac{df}{d\phi}+2f(\phi)\ddot{\phi}\big)\Bigg)$$
\begin{equation}
+\frac{1}{V(\phi)}\Big(\frac{\xi
f(\phi)\dot{\phi}\frac{dV(\phi)}{d\phi}\big(\frac{3}{k^{2}}-9\xi
f(\phi)\dot{\phi}^{2}\big)^{\frac{1}{2}}}{(1-\dot{\phi}^{2})^{\frac{1}{4}}}\Big)
\Big(1-\frac{\dot{\phi}^{2}}{4}+\frac{27\xi
f(\phi)\dot{\phi}^{2}(1-\dot{\phi}^{2})}{2\big(\frac{3}{k^{2}}-9\xi
f(\phi)\dot{\phi}^{2}\big)}\Big)\Bigg].
\end{equation}

This relation is very complicated to find exact analytical
conditions for the stability of the model. So, we perform the
numerical computations to show the evolution of the $C_{s}^{2}$. In
figure 3, we have plotted the $C_{s}^{2}$ for the models considered
in this paper for the numerical calculations shown in figures 1 and
2. From this figure, we can see that the sound speed parameter is
positive
throughout the bouncing phase.\\
\begin{figure}[htp]
\begin{center}
\includegraphics{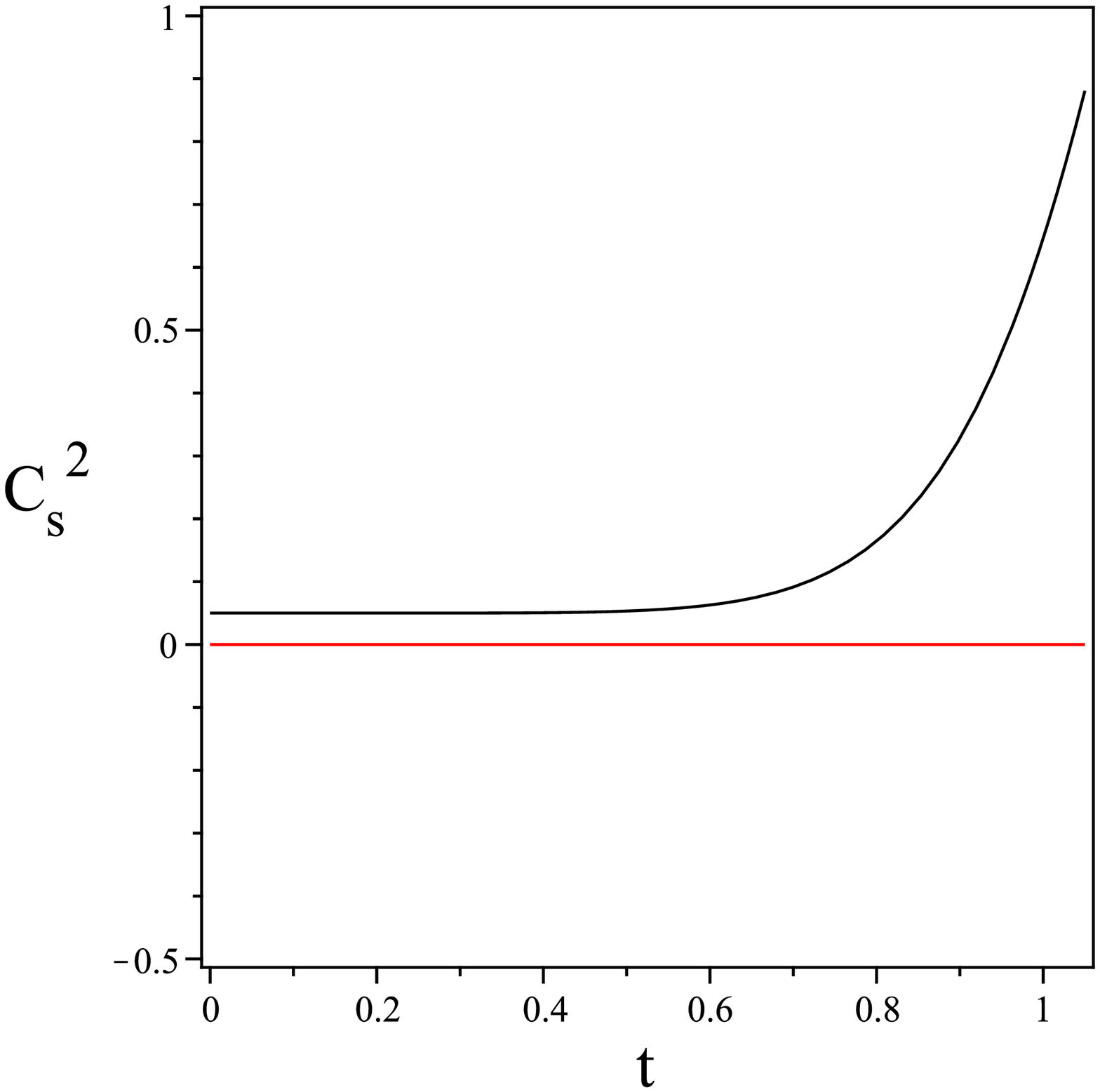} \vspace{8.5cm}\includegraphics{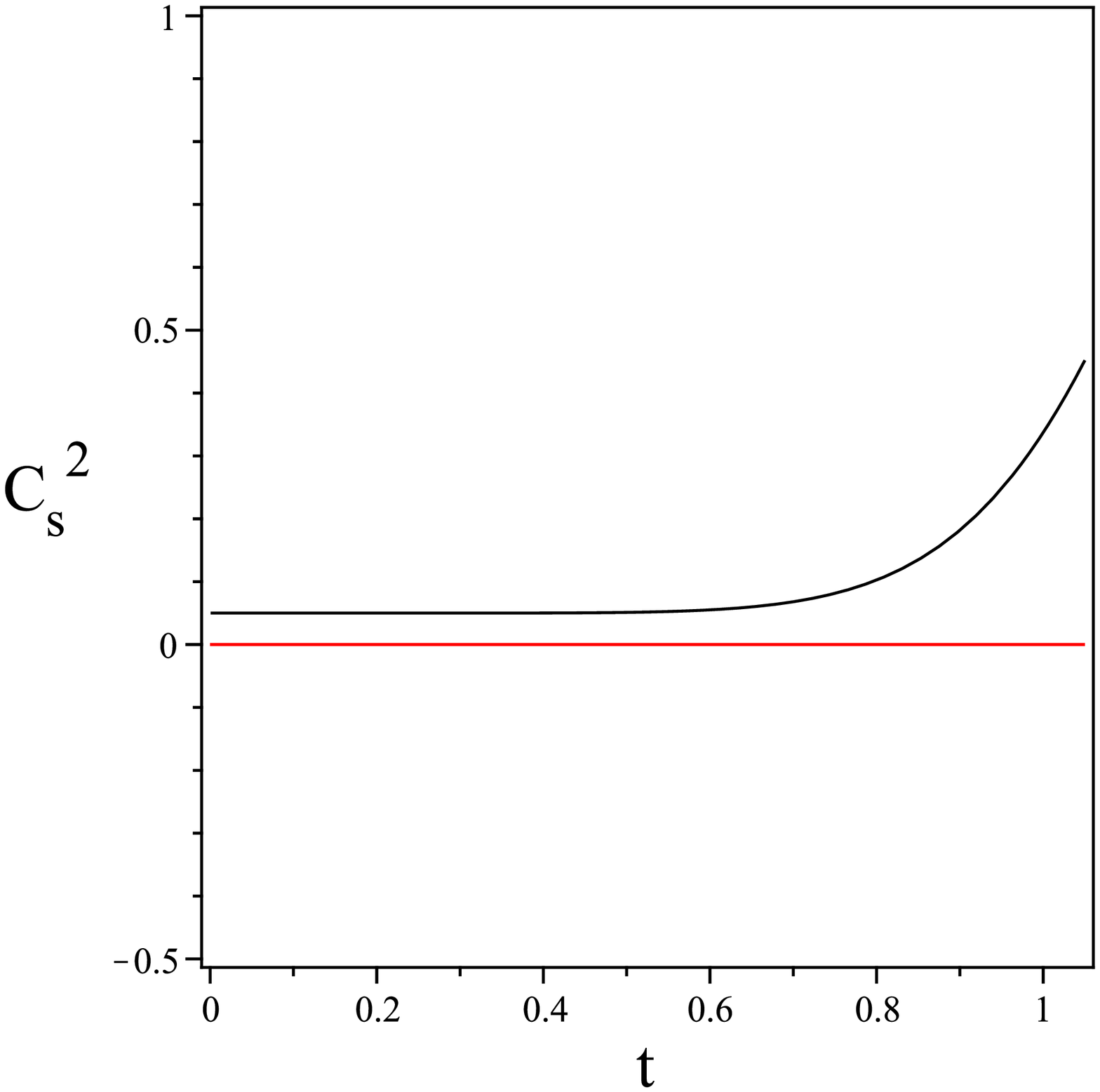}
\end{center}
 \caption{\small { Plots of the sound speeds versus $t$,
 (left for the potential $V(\phi)=V_{0}e^{-\alpha \phi^{2}}$
 and right for the potential $V(\phi)=\frac{V_{0}}{\phi^{2}}$)
, $\phi=\phi_{0}t$ and $f(\phi)=b\phi^{n}$, (with $\xi=10$, $b=1$,
 $n=8$, $V_{0}=4$, $\phi_{0}=0.5$ and $\alpha=5$).}}
\end{figure}

Moreover, we should expect that the coefficient of $\ddot{\phi}$ in
equation (8), is responsible for the presence or absence of ghosts.
We show this coefficient by $D$ (for more details see [61]). In
order to ensure that the perturbation of $\phi$ are not ghosts, one
needs $D>0$, where $D$ is as follows,
\begin{equation}
D=\frac{V(\phi)}{\big(1-\dot{\phi}^{2}\big)^{\frac{3}{2}}}+6\xi
H^{2}f(\phi).
\end{equation}
The evolution of $D$ have been shown in figure 4 for two different
potentials. It is clear from this figure that the function $D$ is
positive during the bouncing phase.
\begin{figure}[htp]
\begin{center}
\includegraphics{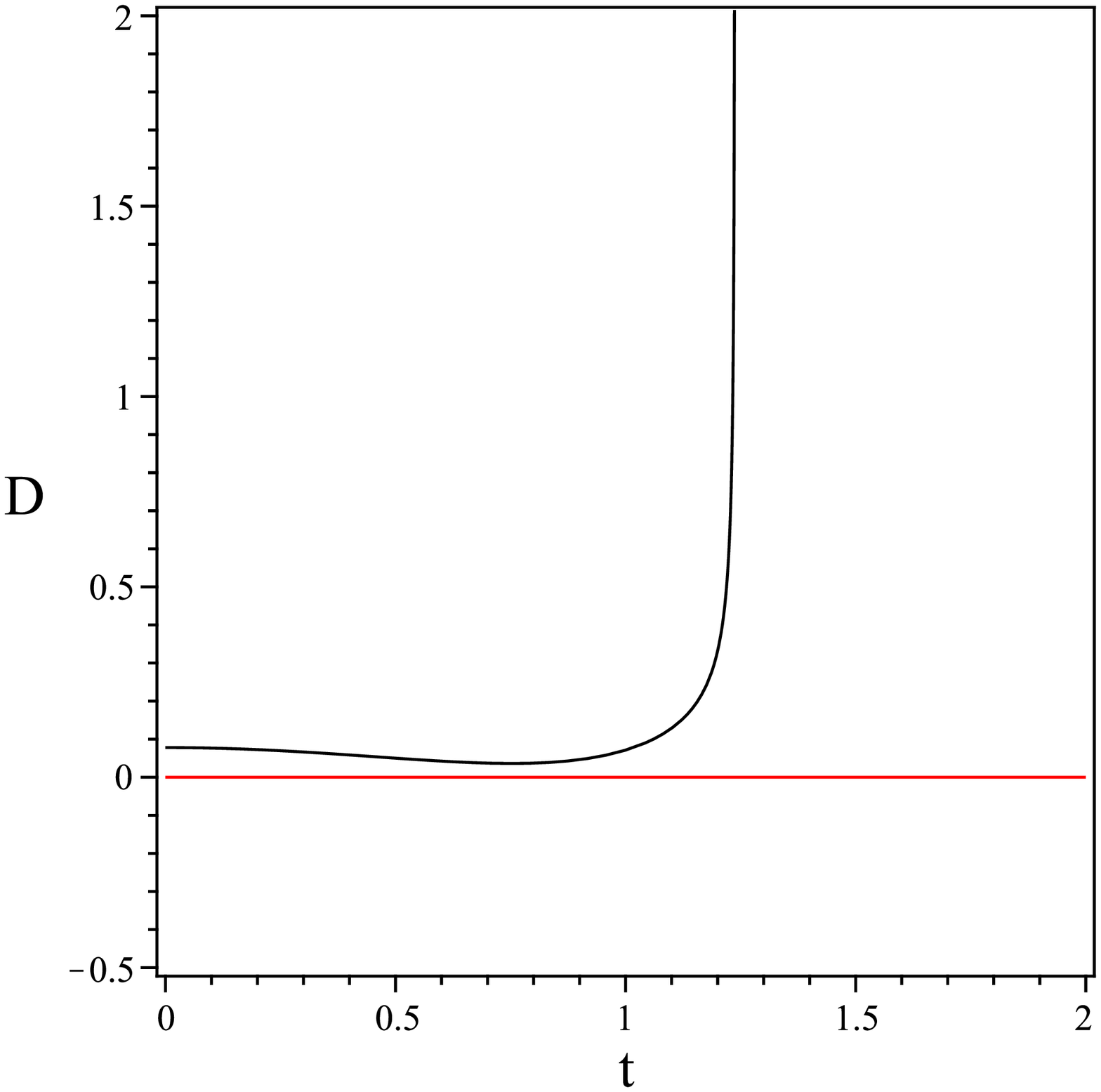} \vspace{8.5cm}\includegraphics{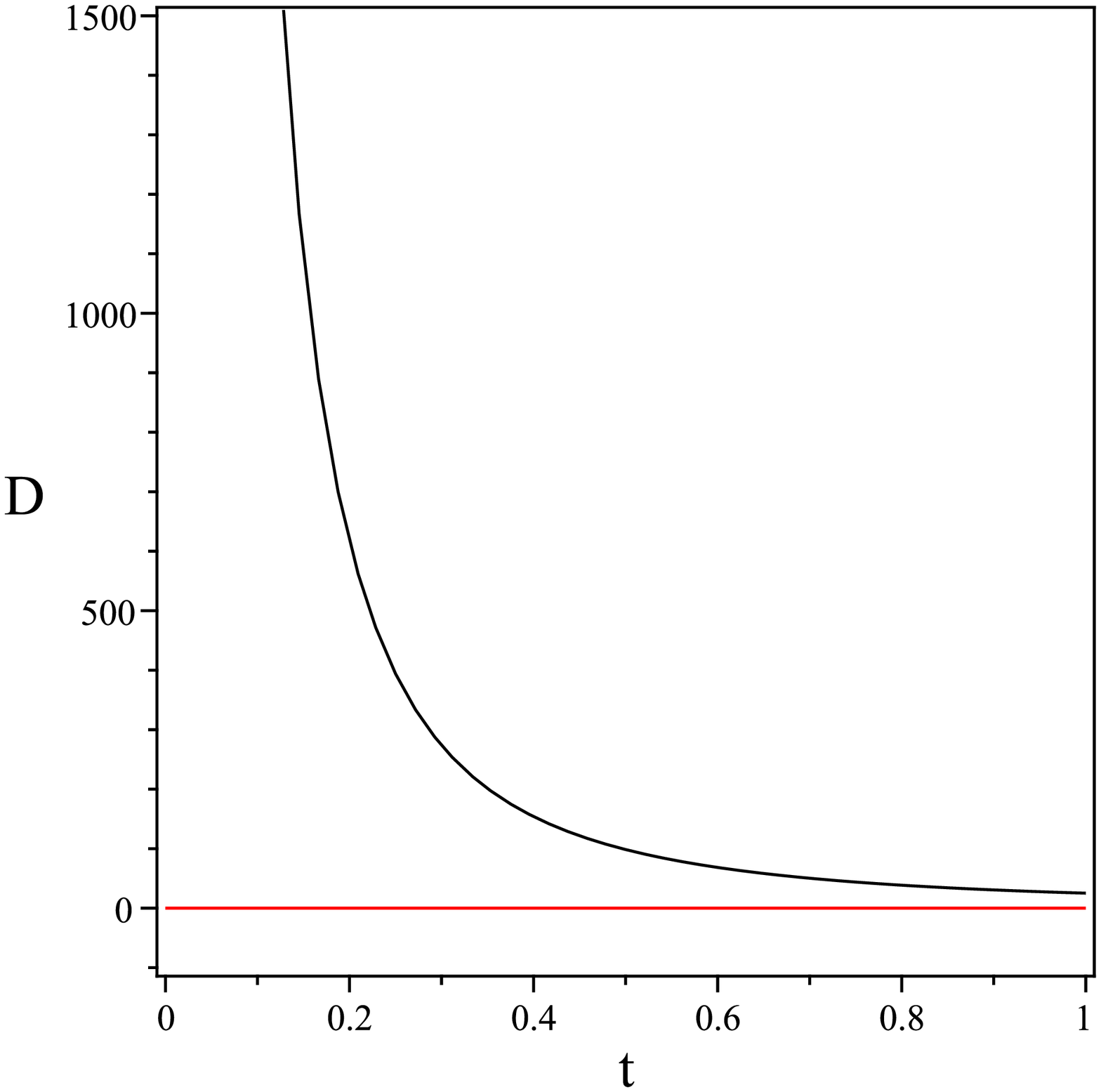}
\end{center}
 \caption{\small { Plots of the $D$ versus $t$,
 (left for the potential $V(\phi)=V_{0}e^{-\alpha \phi^{2}}$
 and right for the potential $V(\phi)=\frac{V_{0}}{\phi^{2}}$)
, $\phi=\phi_{0}t$ and $f(\phi)=b\phi^{n}$, (with $\xi=10$, $b=1$,
 $n=8$, $V_{0}=4$, $\phi_{0}=0.5$ and $\alpha=5$).}}
\end{figure}\\
Now we study the tensor perturbations of our model. We use the
procedure developed in [62] for most general scalar-tensor theories
with second-order field equations. It has been shown that the
quadratic action for the tensor perturbations is as,
\begin{equation}
S_{T}=\frac{1}{8}\int dt\, d^{3}x\,
a^{3}\Big[G_{T}\dot{h}^{2}_{ij}-\frac{F_{T}}{a^{2}}(\vec{\nabla}h_{ij})^{2}\Big],
\end{equation}
$h_{ij}$ is a tensor perturbation satisfying $h_{ij}=0=h_{ij,j}$ and
in case of our model $G_{T}$ and $F_{T}$ are,
\begin{equation}
G_{T}=2\Big(\frac{1}{2\kappa^{2}}-\frac{1}{2}\partial_{\mu}\phi\partial^{\mu}\phi\xi\frac{df(\phi)}{d\phi}\Big),
\end{equation}
\begin{equation}
F_{T}=2\Big(\frac{1}{2\kappa^{2}}+\frac{1}{2}\partial_{\mu}\phi\partial^{\mu}\phi\xi\frac{df(\phi)}{d\phi}\Big).
\end{equation}
In order to avoid ghost and gradient instabilities in the tensor
sector we require the conditions $F_{T}>0$ and $G_{T}>0$.\\
We find the power spectrum of the primordial tensor perturbation as
follows,
\begin{equation}
P_{T}=8\gamma_{T}\frac{G_{T}^{\frac{1}{2}}}{F_{T}^{\frac{3}{2}}}\frac{H^{2}}{4\pi^{2}}\Big|_{-k
y_{T}=1},
\end{equation}
where $d y_{T}=\frac{1}{a}\sqrt{\frac{F_{T}}{G_{T}}}dt$ and $k$ is
the Fourier wavenumber. Note that the sound horizon crossing occurs
when $k^{2}\sim \frac{1}{y_{T}^{2}}$. Also in (21) we have
\begin{equation}
\gamma_{T}=2^{2\nu_{T}-3}\Big|\frac{\Gamma(\nu_{T})}{\Gamma(\frac{3}{2})}\Big|^{2}\big(1-\epsilon-\frac{f_{T}}{2}+\frac{g_{T}}{2}\big),
\end{equation}
here $\nu_{T}=\frac{3-\epsilon+g_{T}}{2-2\epsilon-f_{T}+g_{T}}$,
where
\begin{equation}
g_{T}=\frac{-\dot{\phi}\xi\Big(2\ddot{\phi}\frac{df(\phi)}{d\phi}+\dot{\phi}^{2}\frac{d^{2}f(\phi)}{d\phi^{2}}\Big)}
{2H\Big(\frac{1}{2\kappa^{2}}-\frac{1}{2}\xi\dot{\phi}^{2}\frac{df(\phi)}{d\phi}\Big)},
\end{equation}
\begin{equation}
f_{T}=\frac{\dot{\phi}\xi\Big(2\ddot{\phi}\frac{df(\phi)}{d\phi}+\dot{\phi}^{2}\frac{d^{2}f(\phi)}{d\phi^{2}}\Big)}
{2H\Big(\frac{1}{2\kappa^{2}}+\frac{1}{2}\xi\dot{\phi}^{2}\frac{df(\phi)}{d\phi}\Big)},
\end{equation}
and $\epsilon\equiv -\frac{\dot{H}}{H^{2}}$ that takes the following
form for our model
\begin{equation}
\epsilon=-\frac{\Bigg(9\xi\dot{\phi}\Big(\frac{df(\phi)}{d\phi}\dot{\phi}^{2}+2f(\phi)\ddot{\phi}\Big)\Big(\frac{V(\phi)}{\sqrt{1-\dot{\phi}^{2}
}}+\frac{\Lambda}{\kappa^{2}}\Big)+\frac{\big(\frac{3}{\kappa^{2}}-9\xi
f(\phi)\dot{\phi}^{2}\big)\dot{\phi}}{\sqrt{1-\dot{\phi}^{2}}}\Big(\frac{dV(\phi)}{d\phi}+\frac{2V(\phi)\ddot{\phi}}{(1-\dot{\phi}^{2})}\Big)\Bigg)}
{2 \Big(\frac{V(\phi)}{\sqrt{1-\dot{\phi}^{2}
}}+\frac{\Lambda}{\kappa^{2}}\big)^{\frac{3}{2}}\Big(\frac{3}{\kappa^{2}}-9\xi
f(\phi)\dot{\phi}^{2}\big)^{\frac{1}{2}}}.
\end{equation}
Finally, the tensor spectral index is given by
\begin{equation}
n_{T}=3-2\nu_{T}.
\end{equation}
Thus, one can obtain a blue or red spectrum of tensor perturbations
by choosing different values of coupling parameter $\xi$ in
tachyonic non-minimal derivative coupling model.\\
Let us focus on scalar perturbations by putting $h_{ij}=0$. The
quadratic action for curvature perturbations is as follows,
\begin{equation}
S_{S}=\int dt\, d^{3}x\,
a^{3}\Big[G_{S}\dot{\zeta}^{2}-\frac{F_{S}}{a^{2}}(\vec{\nabla}\zeta)^{2}\Big],
\end{equation}
here $\zeta$ is scalar perturbation in perturbed metric (for more
details see [62]) also $G_{S}$ and $F_{S}$ are given by,
\begin{equation}
G_{S}=\frac{\Xi}{\Theta^{2}}G_{T}^{2}+3G_{T},
\end{equation}
\begin{equation}
F_{S}=\frac{1}{a}\frac{d}{dt}\big(\frac{a}{\Theta}G_{T}^{2}\big)-F_{T},
\end{equation}
where $\Xi$ and $\Theta$ in the case of our model are,
\begin{equation}
\Xi=\frac{\dot{\phi}^{2}V(\phi)}{2(1-\dot{\phi}^{2})^{\frac{3}{2}}}-3H^{2}\Big(\frac{1}{\kappa^{2}}+6\xi
\dot{\phi}^{2}\frac{df(\phi)}{d\phi}\Big),
\end{equation}
\begin{equation}
\Theta=H\Big(\frac{1}{\kappa^{2}}+3\xi
\dot{\phi}^{2}\frac{df(\phi)}{d\phi}\Big).
\end{equation}
By substituting above equations as well as (19) and (20) into
$G_{S}$ and $F_{S}$, one can obtain,
\begin{equation}
G_{S}=\frac{\Big(\frac{1}{\kappa^{2}}-
\xi\dot{\phi}^{2}\frac{df(\phi)}{d\phi}\Big)^{2}}{\Big(\frac{1}{\kappa^{2}}+3\xi
\dot{\phi}^{2}\frac{df(\phi)}{d\phi}\Big)}\Bigg(\frac{\dot{\phi}^{2}V(\phi)}{2H^{2}(1-\dot{\phi}^{2})^{\frac{3}{2}}\Big(\frac{1}{\kappa^{2}}+3\xi
\dot{\phi}^{2}\frac{df(\phi)}{d\phi}\Big)}-3\Bigg)+3\Big(\frac{1}{\kappa^{2}}-
\xi\dot{\phi}^{2}\frac{df(\phi)}{d\phi}\Big),
\end{equation}
$$F_{S}=\Bigg(\frac{1}{\Big(\frac{1}{\kappa^{2}}+3\xi
\dot{\phi}^{2}\frac{df(\phi)}{d\phi}\Big)}-\frac{\Big[\frac{\dot{H}}{\kappa^{2}}+3\xi\Big(\dot{H}\dot{\phi}^{2}\frac{df(\phi)}{d\phi}+2H
\dot{\phi}\ddot{\phi}\frac{df(\phi)}{d\phi}+H\dot{\phi}^{3}\frac{d^{2}f(\phi)}{d\phi^{2}}\Big)\Big]}{H^{2}\Big(\frac{1}{\kappa^{2}}+3\xi
\dot{\phi}^{2}\frac{df(\phi)}{d\phi}\Big)^{2}}\Bigg)\Big(\frac{1}{\kappa^{2}}-\xi\dot{\phi}^{2}\frac{df(\phi)}{d\phi}\Big)^{2}$$
\begin{equation}
+\frac{-2\dot{\phi}\xi}{H\Big(\frac{1}{\kappa^{2}}+3\xi
\dot{\phi}^{2}\frac{df(\phi)}{d\phi}\Big)}\Big(\frac{1}{\kappa^{2}}-\xi\dot{\phi}^{2}\frac{df(\phi)}{d\phi}\Big)
\Big(2\ddot{\phi}\frac{df(\phi)}{d\phi}+\dot{\phi}^{2}\frac{d^{2}f(\phi)}{d\phi^{2}}\Big)-\Big(\frac{1}{\kappa^{2}}+
\xi\dot{\phi}^{2}\frac{df(\phi)}{d\phi}\Big).
\end{equation}
As it is mentioned in [62], the analysis of the curvature
perturbation hereafter is completely parallel to that of the tensor
perturbation. So, ghost and gradient instabilities are avoided as
long as $F_{S}>0$ and $G_{S}>0$.\\
The power spectrum of the primordial curvature perturbation is given
by,
\begin{equation}
P_{\zeta}=\frac{\gamma_{S}}{2}\frac{G_{S}^{\frac{1}{2}}}{F_{S}^{\frac{3}{2}}}\frac{H^{2}}{4\pi^{2}}\Big|_{-k
y_{S}=1},
\end{equation}
where $d y_{S}=\frac{1}{a}\sqrt{\frac{F_{S}}{G_{S}}}dt$ and $k$ is
the Fourier wavenumber. Note that the sound horizon crossing occurs
when $k^{2}\sim \frac{1}{y_{S}^{2}}$. Also in (34) we have
\begin{equation}
\gamma_{S}=2^{2\nu_{S}-3}\Big|\frac{\Gamma(\nu_{S})}{\Gamma(\frac{3}{2})}\Big|^{2}\big(1-\epsilon-\frac{f_{S}}{2}+\frac{g_{S}}{2}\big),
\end{equation}
here we have defined
$\nu_{S}=\frac{3-\epsilon+g_{S}}{2-2\epsilon-f_{S}+g_{S}}$, where
\begin{equation}
g_{S}=\frac{\dot{G_{S}}}{H G_{S}},
\end{equation}
\begin{equation}
f_{S}=\frac{\dot{F_{S}}}{H F_{S}}.
\end{equation}
The spectral index is,
\begin{equation}
n_{S}-1=3-2\nu_{S}.
\end{equation}
The tensor-to-scalar ratio is given by,
\begin{equation}
r=16
\Big(\frac{F_{S}}{F_{T}}\Big)^{\frac{3}{2}}\Big(\frac{G_{S}}{G_{T}}\Big)^{-\frac{1}{2}}.
\end{equation}
\section{Conclusion}
In this work we investigated the bouncing solution in the universe
dominated by the quintom matter. We analyzed the possibility of
obtaining a cosmological bounce in a model which contains a scalar
field with non-minimal derivative coupling to Einstein tensor and
itself too. We used the numerical methods to study the bouncing
behavior in this setup where the tachyon field played the role of
scalar field. We obtained the bouncing condition as equation (13),
then by considering a couple example for potential of the tachyon
scalar field we showed that the above mentioned condition can be
satisfied. After that we considered the conditions required for the
classical stability of our model. These conditions obtained as
equations (16) and (17) which reflect the positivity of the sound
speed $C_{s}^{2}$ and $D$. Our numerical results for such a
classical stability plotted in figures 3 and 4 respectively.
Finally, we investigated the tensor perturbations of our model using
the method of Ref. [62].\\
\\
\textbf{Acknowledgements} \\
The authors are indebted to the anonymous referee for his/her
comments that improved the paper drastically.


\begin{thebibliography}{11}
\bibitem{c1}
D. N. Spergel et al. [WMAP Collaboration], {\it Astrophys. J.
Suppl.} {\bf 148}, 175 (2003).
\bibitem{c2}
A. Linde, {\it Particle Physics and Inflationary Cosmology}
(Harwood, Chur, Switzerland, 1990).
\bibitem{c3}
R. C. Tolman, {\it Phys. Rev.} {\bf 37}, 1639 (1931).
\bibitem{c4}
V. F. Mukhanov and R. H. Brandenberger, {\it Phys. Rev. Lett.} {\bf
68}, 1969 (1992); R. H. Brandenberger, V. F. Mukhanov and A. Sorn-
borger, {\it Phys. Rev. D} {\bf 48}, 1629 (1993).
\bibitem{c5}
M. Novello and S. E. P. Bergliaffa, {\it Phys. Rept.} {\bf 463}, 127
(2008).
\bibitem{c6}
A. Borde and A. Vilenkin, {\it Phys. Rev. Lett.} {\bf 72}, 3305
(1994).
\bibitem{c7}
G. Veneziano, {\it Phys. Lett. B} {\bf 265}, 287 (1991); M.
Gasperini and G. Veneziano, {\it Astropart. Phys.} {\bf 1}, 317
(1993).
\bibitem{c8}
J. Khoury, B. A. Ovrut, P. J. Steinhardt and N. Turok, {\it Phys.
Rev. D} {\bf 64}, 123522 (2001); J. Khoury, B. A. Ovrut, N. Seiberg,
P. J. Steinhardt and N. Turok, {\it Phys. Rev. D} {\bf 65}, 086007
(2002).
\bibitem{c9}
Y. F. Cai, S. H. Chen, J. B. Dent, S. Dutta and E. N. Saridakis,
[astro-ph.CO/1104.4349].
\bibitem{c10}
R. Brustein and R. Madden, {\it Phys. Rev. D} {\bf 57}, 712 (1998);
T. Biswas, A. Mazumdar and W. Siegel, {\it JCAP} {\bf 0603}, 009
(2006); T. Biswas, R. Brandenberger, A. Mazumdar and W. Siegel, {\it
JCAP} {\bf 0712}, 011 (2007).
\bibitem{c11}
R. Brandenberger, {\it Phys. Rev. D} {\bf 80}, 043516 (2009).
\bibitem{c12}
 Y. F. Cai and E. N. Saridakis, {\it JCAP} {\bf 0910}, 020 (2009);
 E. N. Saridakis, {\it Eur. Phys. J. C} {\bf 67}, 229 (2010).
\bibitem{c13}
M. Bojowald, {\it Phys. Rev. Lett.} {\bf 86}, 5227 (2001).
\bibitem{c14}
R. H. Brandenberger and C. Vafa, {\it Nucl. Phys. B} {\bf 316}, 391
(1989).
\bibitem{c15}
R. H. Brandenberger, [hep-th/0808.0746].
\bibitem{c16}
T. Biswas, R. Brandenberger, A. Mazumdar and W. Siegel, {\it JCAP}
{\bf 0712}, 011 (2007).
\bibitem{c17}
V. Bozza and G. Veneziano, {\it Phys. Lett. B} {\bf 625}, 177
(2005).
\bibitem{c18}
P. Peter and N. Pinto-Neto, {\it Phys. Rev. D} {\bf 66}, 063509
(2002).
\bibitem{c19}
M. R. Setare, J. Sadeghi and A. Banijamali, {\it Phys. Lett. B} {\bf
669}, 9 (2008); T. Qiu and K. C. Yang, {\it JCAP} {\bf 1011}, 012
(2010); T. Qiu, {\it Class. Quant. Grav.} {\bf 27}, 215013 (2010).
\bibitem{c20}
E. I. Buchbinder, J. Khoury and B. A. Ovrut, {\it Phys. Rev. D} {\bf
76}, 123503 (2007).
\bibitem{c21}
P. Creminelli and L. Senatore, {\it JCAP} {\bf 0711}, 010 (2007).
\bibitem{c22}
C. Lin, R. H. Brandenberger and L. P. Levasseur, {\it JCAP} {\bf
1104}, 019 (2011).
\bibitem{c23}
A. A. Starobinsky, {\it Sov. Astron. Lett.} {\bf 4}, 82 (1978).
\bibitem{c24}
Y. F. Cai, T. Qiu, Y. S. Piao, M. Li and X. Zhang, {\it JHEP} {\bf
0710}, 071 (2007); Y. F. Cai, T. Qiu, R. Brandenberger, Y. S. Piao
and X. Zhang, {\it JCAP} {\bf 0803}, 013 (2008);Y. F. Cai, T. T.
Qiu, R. Brandenberger and X. m. Zhang, {\it Phys. Rev. D} {\bf 80},
023511 (2009).
\bibitem{c25}
Y. F. Cai, T. t. Qiu, J. Q. Xia and X. Zhang, {\it Phys. Rev. D}
{\bf 79}, 021303 (2009); Y. F. Cai and X. Zhang, {\it JCAP} {\bf
0906}, 003 (2009); J. Liu, Y. F. Cai and H. Li,
[astro-ph.CO/1009.3372].
\bibitem{c26}
S. Nesseris and L. Perivolaropoulos, {\it J. Cosmol. Astropart.
Phys.} {\bf 01}, 018 (2007); U. Alam, V. Sahni and A.A. Starobinsky,
{\it JCAP} {\bf 0406}, 008 (2004); H. K. Jassal, J. S. Bagla, and T.
Padmanabhan, {\it Mon. Not. Roy. Astron. Soc.} {\bf 405}, 2639
(2010).
\bibitem{c27}
B. Boisseau, G. Esposito-Farese, D. Polarski and A.A. Starobinsky,
{\it Phys. Rev. Lett.} {\bf85}, 2236 (2000).
\bibitem{c28}
B. Feng, X. Wang and X. Zhang, {\it Phys. Lett. B} {\bf 607}, 35
(2005).
\bibitem{c29}
S. Nojiri, S. D. Odintsov and M. Sasaki, {\it Phys. Rev. D} {\bf
71}, 123509 (2005); M. Sami, A. Toporensky, P. V. Tretjakov and S.
Tsujikawa, {\it Phys. Lett. B} {\bf 619}, 193 (2005); B. M. Leith
and I. P. Neupane, {\it JCAP} {\bf 0705}, 019 (2007); S. Nojiri, S.
D. Odintsov and M. Sami, {\it Phys. Rev. D} {\bf 74}, 046004 (2006);
T. Koivisto and D. F. Mota, {\it Phys. Lett. B} {\bf 644}, 104
(2007); M. R. Setare and E. N. Saridakis, {\it Phys. Lett. B} {\bf
670}, 1 (2008); A. K. Sanyal, [astro-ph/0710.2450].
\bibitem{c30}
A. Vikman, {\it Phys. Rev. D} {\bf 71}, 023515 (2005).
\bibitem{c31}
C. Deffayet, G. R. Dvali and G. Gabadadze, {\it Phys. Rev. D} {\bf
65}, 044023 (2002);  M. R. Setare, {\it Phys. Lett. B} {\bf 642},
421, (2006).
\bibitem{c32}
M. Z. Li, B. Feng and X. M. Zhang, {\it JCAP} {\bf 0512}, 002
(2005).
\bibitem{c33}
M. Cataldo and L. P. Chimento, [astro-ph/0710.4306].
\bibitem{c34}
B. McInnes, {\it Nucl. Phys. B} {\bf 718}, 55 (2005); R. G. Cai, H.
S. Zhang and A. Wang, Commun. {\it Theor. Phys.} {\bf 44}, 948
(2005); R. G. Cai, Y. g. Gong and B. Wang, {\it JCAP} {\bf 0603},
006 (2006); I. Y. Aref'eva, A. S. Koshelev and S. Y. Vernov, {\it
Phys. Rev. D} {\bf 72}, 064017 (2005); G. Kofinas, G. Panotopoulos
and T. N. Tomaras, {\it JHEP} {\bf 0601}, 107 (2006); I. Y. Aref'eva
and A. S. Koshelev, {\it ibid.} {\bf 0702}, 041 (2007); L. P.
Chimento, R. Lazkoz, R. Maartens and I. Quiros, {\it JCAP} {\bf
0609}, 004 (2006); P. S. Apostolopoulos and N. Tetradis, {\it Phys.
Rev. D} {\bf 74}, 064021 (2006); S. F. Wu, A. Chatrabhuti, G. H.
Yang and P. M. Zhang, {\it ibid.} {\bf 659}, 45 (2008); J. Sadeghi,
M. R. Setare, A. Banijamali and F. Milani, {\it Phys. Lett. B} {\bf
662}, 92 (2008).
\bibitem{c35}
K. Bamba, C. -Q. Geng, S. Nojiri and S. D. Odintsov, {\it Phys. Rev.
D} {\bf 79}, 083014 (2009).
\bibitem{c36}
R. R. Caldwell, {\it Phys. Lett. B} {\bf 545}, 23 (2002); S. Nojiri
and S. D. Odintsov, {\it Phys. Lett. B} {\bf 562}, 147 (2003).
\bibitem{c37}
L. Perivolaropoulos, {\it JCAP} {\bf 0510}, 001 (2005).
\bibitem{c38}
J. Sadeghi, M. R. Setare, A. Banijamali and F. Milani, {\it Phys.
Rev. D} {\bf 79}, 123003 (2009).
\bibitem{c39}
T. Padmanabhan, {\it Phys. Rept.} {\bf 380}, 235 (2003).
\bibitem{c40}
C. Deffayet, O. Pujolas, I. Sawicki and A. Vikman, {\it JCAP} {\bf
1010}, 026 (2010); O. Pujolas, I. Sawicki and A. Vikman,
[hep-th/1103.5360].
\bibitem{c41}
E. A. Lim, I. Sawicki and A. Vikman, {\it JCAP} {\bf 1005}, 012
(2010).
\bibitem{c42}
V. Faraoni, {\it Phys. Rev. D} {\bf 62}, 023504 (2000).
\bibitem{c43}
L. Amendola, {\it Phys. Lett. B} {\bf 301}, 175 (1993).
\bibitem{c44}
S. Capozziello and G. Lambiase, {\it Gen. Rel. Grav.} {\bf 31}, 1005
(1999).
\bibitem{c45}
S. Capozziello, G. Lambiase and H.-J.Schmidt, {\it Annalen Phys.}
{\bf 9}, 39 (2000).
\bibitem{c46}
C. Germani and A. Kehagias, [hep-ph/1003.2635].
\bibitem{c47}
S. V. Sushkov, {\it Phys. Rev. D} {\bf 80}, 103505 (2009).
\bibitem{c48}
E. N. Saridakis and S. V. Sushkov, {\it Phys. Rev. D} {\bf 81},
083510 (2010); H. M. Sadjadi, {\it Phys. Rev. D} {\bf 83} 107301
(2011).
\bibitem{c49}
S. F. Daniel and R. R. Caldwell, {\it Class. Quant. Grav} {\bf 24},
5573 (2007).
\bibitem{c50}
L. N. Granda and W. Cardona, {\it JCAP} {\bf 07}, 021 (2010).
\bibitem{c51}
S. Nojiri, S. D. Odintsov and P. V. Tretyakov, {\it
Prog.Theor.Phys.Suppl.} {\bf 172}, 81-89 (2008).
\bibitem{c52}
S. Alexander, {\it Phys. Rev. D} {\bf 65}, 023507 (2002); A.
Mazumdar, S. Panda and A. Perez-Lorenzana, {\it Nucl. Phys. B} {\bf
614}, 101 (2001); G. Gibbons, {\it Phys. Lett. B} {\bf 537}, 1
(2002).
\bibitem{c53}
A. Sen, {\it JHEP} {\bf 9910}, 008 (1999); E. Bergshoeff, M. de Roo,
T. de Wit, E. Eyras and S. Panda, {\it JHEP} {\bf 0005}, 009 (2000);
J. Kluson, {\it Phys. Rev. D} {\bf 62}, 126003 (2000).
\bibitem{c54}
M. B. Green, J. H. Schwarz and E. Witten, {\it Superstring Theory},
Cambridge University Press (1987).
\bibitem{c55}
H. Liu and A.A. Tseytlin, {\it Nucl. Phys. B} {\bf 533}, 88 (1998).
\bibitem{c56}
S. Nojiri and S. D. Odintsov, {\it Phys. Lett. B} {\bf 444}, 92
(1998).
\bibitem{c57}
Q. Shafi and C. Wetterich, {\it Phys. Lett. B} {\bf 152}, 51 (1985).
\bibitem{c58}
Q. Shafi and C. Wetterich, {\it Nucl. Phys. B} {\bf 289}, 787
(1987).
\bibitem{c59}
A. Banijamali and B. Fazlpour, {\it Phys. Lett. B} {\bf 703}, 366
(2011).
\bibitem{c60}
L.R. Abramo and N. Pinto-Neto, {\it Phys. Rev. D} {\bf 73}, 063522
(2006).
\bibitem{c61}
O. Pujolas, I. Sawicki and A. Vikman, [hep-th/1103.5360v2].
\bibitem{c62}
T. Kobayashi, M. Yamaguchi and J. Yokoyama, {\it Prog. Theor. Phys.}
{\bf 126}, 511 (2011).
\end{thebibliography}
\end{document}